\documentclass{aa}  

\usepackage{graphicx}
\graphicspath{{./Plots/}}
\usepackage[dvipsnames]{xcolor}
\usepackage{float}
\usepackage{amsmath}
\usepackage{longtable}
\usepackage{gensymb}
\usepackage{listings}
\usepackage{mathtools}
\usepackage{amssymb}
\usepackage{fancyhdr}
\usepackage{caption}
\usepackage{multirow}
\usepackage{tabularx}
\usepackage{subcaption}
\usepackage{txfonts}

\usepackage[colorlinks=true,
            linkcolor=MidnightBlue,
            citecolor=MidnightBlue,
            urlcolor=MidnightBlue
            ]{hyperref}   

\newcommand{\twave}{\tau_{\rm wave}}
\newcommand{\Msun}{\mathrm{M}_\odot}
\newcommand{\Mearth}{\mathrm{M}_\oplus}
\newcommand{\mpl}{m_\mathrm{p}}
\renewcommand{\epsilon}{\varepsilon}
\newcommand{\ep}{\epsilon_\mathrm{p}}
\newcommand{\Siggap}{\Sigma_{0,\rm{gap}}}

\begin{document}

   \title{Mean motion resonance capture in the context of type-I migration}

   \author{Kaltrina Kajtazi\inst{1,}, Antoine C. Petit\inst{1,2}
          \and
          Anders Johansen \inst{1,2}
          }

   \institute{Lund Observatory, Department of Astronomy and Theoretical Physics, Lund University, Box 43, 22100 Lund, Sweden
   \and
    Center for  Star and  Planet  Formation,  GLOBE  Institute,  University  of  Copenhagen,  ØsterVoldgade 5-7, 1350 Copenhagen, Denmark\\
        \email{1kaltrinakajtazi@gmail.com}}
   \date{Received , ; accepted , }

\abstract{Capture into mean motion resonance (MMR) is an important dynamical mechanism as it shapes the final architecture of a planetary system.
We simulate systems of two or three planets undergoing migration with varied initial parameters such as planetary mass and disk surface density and analyse the resulting resonant chains. 
In contrast to previous studies, our results show that the disk properties have the dominant impact on capture into mean motion resonance, while the total planetary mass barely affects the final system configuration as long as the planet does not open a gap in the disk.
We confirm that the adiabatic resonant capture is the correct framework to understand the conditions leading to MMR formation, since its predictions are qualitatively similar to the numerical results.
However, we find that the eccentricity damping can facilitate the capture in a given resonance.
We find that under typical disk conditions, planets tend to be captured into 2:1 or 3:2 MMRs, which agrees well with the observed exoplanet MMRs.
Our results predict two categories of systems: those that have uniform chains of wide resonances (2:1 or 3:2 MMRs) and those that have a more compact inner pair than the outer pair such as 4:3:2 chains.
Both categories of resonant chains are present in observed exoplanet systems. 
On the other hand, chains with a wider inner pair than the outer one are very rare and emerge from stochastic capture.
Our work here can be used to link current configuration of exoplanetary systems to the formation conditions within protoplanetary disks.
}


   \maketitle

\section{Introduction}\label{sec:intro}

Planets with a mass comprised between Earth's and Neptune's (called super-Earths) orbiting their star on orbits shorter than 100 days are the most common exoplanets discovered.
It is estimated that close to 30\% of Sun-like stars host a super-Earth \citep{Fressin2013,Petigura2018}.
Among them, many are in multiplanetary systems (see \citealp{Zhu2021} for a review).
As planets grow in the protoplanetary disk, they raise spiral density wakes which exert a gravitational torque, leading to their migration \citep{Lin1979,Goldreich1980,Ward1997,Armitage2020}.
In this regime, known as type-I migration, a super-Earth spirals in at a rate of the order of $1\, \mathrm{m\, s^{-1}}$ until it reaches the inner edge of the disk where a positive torque stops the migration \citep{Masset2006}.
Hydrodynamical 2D and 3D simulations of protoplanets embedded in a disk have constrained the migration rate as a function of the disk and planet properties \citep{Tanaka2002,Tanaka2004,Cresswell2006,Cresswell2008,Baruteau2014,McNally2020}.
The migration path depends on the properties of the disk and the planet. 
A steep temperature gradient can slow or even reverse migration \citep{Paardekooper2010,Bitsch2015}, or the accretion heat from the growing planet can modify the corotation torques \citep{Benitez-Llambay2015}.
If the planet becomes more massive than a a few tens of Earth masses, it will open a gap within the protoplanetary disk, slowing or even halting the migration \citep[\emph{e.g.}][]{DAngelo2010,Lega2021}.
This regime, called type-II migration has recently been interpreted as an extension of the type-I migration by accounting for the reduced gas surface density within the gap \citep{Kanagawa2018}.
But in all cases, planet migration is a natural and an unavoidable mechanism in the context of formation within the protoplanetary disk.
Combined with the planet growth through pebble accretion \citep{Lambrechts2012,Johansen2017}, planet migration is a dominant physical process to shape the architecture of planetary systems \citep{Johansen2019}.

When multiple planets grow and migrate towards the inner disk edge, their gravitational interactions lead to capture into orbital resonances \citep{Peale1976,Goldreich1980}.
Mean motion resonances (MMR) are configurations where the planet period ratios are close to a ratio of integers.
While the majority of close-in multiplanetary systems observed with the \emph{Kepler} satellite are out of MMRs \citep{Fabrycky2014}, the existence of more than a hundred systems close to first-order MMRs (that is MMRs such that the period ratio is close to $(k+1)/k$) is an important evidence that migration occurs during the planet formation.
Long MMR chains are a common feature in state-of-the-art planet formation simulations \citep{Izidoro2017,Lambrechts2019,Izidoro2021} where many systems emerge from the disk phase in such a configuration before being broken soon after the disk dispersal.
\cite{Izidoro2017} estimate that roughly 95\% of the resonant chain should be broken to be consistent with the observations.
Many mechanisms have been proposed to break resonant chains such as intrinsic instability \citep{Matsumoto2012,Pichierri2020}, turbulence within the disk \citep{Batygin2017}, a change in the planet or stellar mass \citep{Matsumoto2020} or tidal dissipation \citep{Delisle2014}, possibly enhanced by chaotic obliquities \citep{Millholland2019}.
Observed resonant chains provide a unique opportunity to probe the architecture of pristine configurations, before orbital rearrangement.
Studies on specific systems can offer a window on their formation conditions \citep{Teyssandier2020, Petit2020,Huhn2021,Huang2022}, or on the resonance dynamics and the capture mechanism \citep{Pichierri2019,Charalambous2022}.

Understanding the resonance capture has been the subject of numerous works over the past decades (see \citealp{Batygin2015} for a complete review).
After the first studies \citep{Goldreich1965,Goldreich1966}, a first analytical criterion was proposed in the context of the circular restricted three-body problem developing the concept of probabilistic capture \citep{Yoder1979}.
The first general theory was developed using the formalism of adiabatic invariant by \cite{Henrard1982} (see also \citealp{Henrard1983,Lemaitre1984}).
\cite{Batygin2015} has since simplified the framework and applied it directly to the context of capture of two massive planets undergoing migration.
Capture is only possible for convergent migration, at a sufficiently slow pace.
If the migration is slow enough to meet the adiabatic threshold (see section \ref{sec:theory}), the capture is guaranteed for small eccentricities and has a finite probability for eccentric orbits.
\cite{Mustill2011} and \citet{Ogihara2013} have numerically studied how the migration parameters affect the boundaries for capture and its probabilities.
Other studies have focused on extending the capture theory to the non-adiabatic regime \citep{Friedland2001,Quillen2006,Ketchum2011}.
After the resonance capture, a system can nevertheless escape the resonance due to the overstability of the fixed point \citep{Papaloizou2010,Deck2015,Xu2018} or be dynamically unstable \citep{Pichierri2018}.
Moreover, systems in MMR can also be affected by general dynamical instabilities such as MMR overlap \citep{Wisdom1980,Deck2013,Petit2017,Hadden2018,Petit2020a}.

While many works have focused on the understanding of resonance capture, there is still a need for a general study of the capture in first-order MMR under a realistic set-up.
Most other numerical studies focus on the capture into a specific resonance and initialize the systems just outside of said resonance without a study on the fate of a system if it were to cross the MMR.
Besides that, convergent type-I migration for a pair of migrating planets is only possible if the outer planet is more massive or if the inner planet is halted by a planet trap.
However, a configuration with a more massive outer planet often leads to an unstable resonance \citep{Deck2013,Xu2018}, and the case of a trapped inner planet has not been studied in details, despite the fact that such configuration are supported by theoretical models (planet growth tracks \citealp{Johansen2019}) and observations (uniform "peas in a pod" systems \citealp{Weiss2018}).

Our goal is to investigate the parameters governing migration and resonant capture to see which MMRs are created as a function of the initial planetary and disk properties.
While the typical conditions in the protoplanetary disk, as well as the planetary masses, can be roughly constrained, 
we explicitly probe how capture in resonance changes over a large set of parameters covering the regime where planets experience type-I migration as well as the limits of this regime where planets start to open a gap.
We analyse our results in the type-I regime by considering the dynamically relevant variables (see section \ref{sec:theory}), in order to compare our results to analytical criteria.
Our study probe disk setups extending beyond the physically motivated conditions in order to contextualise the observational constraints.
Indeed, if the capture outcomes are widely different for unrealistic conditions in the protoplanetary disk and the planetary system (such as very large gas surface densities), we can use our results to consolidate our interpretation of the observed resonant chains.
Thereby, our work can be viewed as a generalization of the numerical works of \cite{Ogihara2013} and \cite{Xu2018}, that studied the capture behaviour as a function of the migration prescriptions in a limited context.

In Section \ref{sec:method}, we detail our model, the numerical setup and the previously proposed MMR capture analytical criteria.
In Section \ref{sec:results}, we present the results from our numerical experiments on two equal mass planet systems for various masses, surface density and disk aspect ratio.
We then generalise our results by considering three equal mass planet systems in section \ref{sec:3p}.
We investigate in Section \ref{sec:s0_k},  the trends in capture condition between the different first-order MMRs by combining our previous results.
Finally, in Section \ref{sec:discussion} we compare our findings to exoplanet resonant systems and discuss the theoretical implications of our study.

\section{Numerical setup and theoretical motivation}\label{sec:method}

\subsection{Setup conditions}\label{subs:setup}

We use the N-body numerical integration library \texttt{Rebound}, with the integrator \texttt{WHfast} \citep{Rein2012,Rein2015} to perform our simulations. 
We integrate coplanar systems with a $1\, \Msun$ star, in a non-evolving protoplanetary disk of surface density $\Sigma$ and scale height $h$ defined as
\begin{equation}
    \Sigma = \Sigma_0 \left(\frac{r}{1\, {\rm AU}}\right)^{-s},
    \label{eq:sigma}
\end{equation}
\begin{equation}
    h = h_0\left(\frac{r}{1\, {\rm AU}}\right)^{\beta},
    \label{eq:h}
\end{equation}
where $s$ is the surface density profile index, $\beta$ is the flaring index, $\Sigma_0$ is the surface density at $1\, {\rm AU}$ and $h_0$ is the aspect ratio at $1\, {\rm AU}$.
Following \citet{Pichierri2018}, we adopt the same values $s=1$ and $\beta=0.25$, which correspond to an optically thin disk \citep{Hayashi1981} assuming a constant accretion rate and constant $\alpha$-viscosity parameter \citep{Ida2016}.

In the standard parameterized $\alpha$ viscous spreading prescription \citep{Shakura1973}, the gas accretion rate is connected to the surface density and aspect ratio through the relationship
\begin{equation}
    \dot{M}_{\rm g}= 3\pi\alpha h^2\Sigma\Omega_{\rm K}r^2, \label{eq:Mdotgen}
\end{equation}
where $\alpha$ is the viscous coefficient and $\Omega_{\rm K}$ the Keplerian velocity.
In a steady state, the accretion rate is constant in the disk and we can evaluate Eq.~\eqref{eq:Mdotgen} at $r_0=1\, {\rm AU}$
\begin{equation}
    \dot{M}_{\rm g}=  3\pi\alpha h_0^2\Sigma_0\sqrt{\mathcal{G} M_*r_0}\label{eq:Mdot},
\end{equation}
where $\mathcal{G}$ is the gravitational constant.
The accretion rate is an observable which allows us to constrain the gas surface density that is still poorly constrained in protoplanetary disks.
Mass accretion rate observations and inferred gas surface densities are consistent with $\alpha$ in the broad range $10^{-3}$--$10^{-2}$ \citep{Hartmann1998,Andrews2009}.
Within the disk lifetime, the accretion rate declines from $10^{-7}\, \mathrm{M_\odot\, yr^{-1}}$ to $10^{-9}\, \mathrm{M_\odot\, yr^{-1}}$ \citep{Manara2016,Tazzari2017,Manara2022}.
Importantly, the accretion rate can be loosely connected to the system age during the formation phase.
We however emphasize that we only infer a disk accretion rate to connect our results to observable quantities.
In particular, the assumed value for $\alpha$ does not affect the results of our simulations, as the disk is parameterized through its surface density and aspect ratio.

We consider a case where the inner planet is fixed at $0.1\, {\rm AU}$ at the inner disk edge, while the outer one migrates inward via type-I migration, from a position outside the 2:1 MMR with the inner planet. 
The initial positions of the planets are $0.1\, {\rm AU}$ and  $0.2\, {\rm AU}$ unless specified otherwise. 
Fixing the inner planet at the inner disk edge provides a controlled setting where the planets are captured into MMR at the same position and orbital period. 
Based on planet formation models, we can consider the migration to be sequential, meaning planets grow then migrate until they reach the inner disk edge one after the other \citep{Izidoro2017}, 
possibly entering and maintaining a mean motion resonant configuration. 

Every integration lasts for one migration timescale of the outer planet. This is more than enough time for the outer planet to be captured in a MMR configuration as the planet would reach the inner disk edge if it were isolated.
We do not integrate for longer since we do not focus on the long term stability of the MMR. 
We choose a timestep of $P\rm (0.1\, AU)/20 = 1.6\times 10^{-3}\, {\rm yr}$, which is sufficient for this kind of setup. 
Moreover, we stop the integration if the separation between the planets is smaller than one Hill radius. 
Smaller separations result in encounters or collisions and will not lead to a stable MMR configuration. 
Initial eccentricity and mutual inclinations are kept unchanged at zero as they are rapidly damped by the disk and do not affect MMR capture \citep{Mustill2011,Ogihara2013}.

\subsection{Planet migration}\label{subs:planet_mig}

Within the protoplanetary disk, a protoplanet exchanges angular momentum with the gas and migrates \citep[\emph{e.g.}][]{Armitage2020}.
If the perturbation to the gas density remains small and the planet does not open a gap, the mechanism is called type-I migration \citep{Tanaka2002,Cresswell2006,Baruteau2014}.
The disk-planet interaction is composed of two separate contributions, damping the orbit's eccentricity and the migration affecting the semi-major axis.
More precisely, the eccentricity is dampened as
\begin{equation}
    \dot{e}_{{\rm damp}} = -\frac{e}{\tau_{e}}.
    \label{eq:edamp}
\end{equation}
and the disk interaction's contribution to the semi-major axis dynamics is
\begin{equation}
    \dot{a}_{{\rm mig}} = \left(-\frac{1}{\tau_{a}} - 2p\frac{e^2}{\tau_{e}}\right)a,
    \label{eq:adamp}
\end{equation}
where $p = 1$ for small values of $e$ \citep{Pichierri2018}. The timescales $\tau_{a}$ and $\tau_{e}$ are then defined as
\begin{align}
    \tau_{a} =& \frac{\twave h^{-2}}{(2.7+1.1s)}f_{ei,a},\label{eq:ta}  \\
    \tau_{e} =& \frac{\twave}{0.780}f_{ei,e}, \label{eq:te}
\end{align}
where \(\twave\) is the typical timescale of type-I migration \citep{Tanaka2004}
\begin{equation}
    \twave = \frac{M_*^2h^4}{\mpl\Sigma a^2 n},
    \label{eq:twave}
\end{equation}
$f_{ei,a}$ and $f_{ei,e}$ are functions describing the influence of the inclination and eccentricity terms, and $n$ is the planetary mean motion.
In this work, we use the expressions in Eqs. (11-13) of \citet{Cresswell2008}. 
While more realistic torque expressions have been developed since \citep{Paardekooper2010,Paardekooper2011}, they change the prefactor of the migration timescale $\tau_a$ by a factor of order unity which varies slowly with the planet mass.
This change would slightly affect the migration timescales for a given surface density but not the relative migration speed between the planets.
As we consider a wide range of masses and surface densities, changing the torque prescription would only affect moderately the quantitative results.

In the numerical simulations, the effects of damping $e$ and $a$ are applied to the planet acceleration as
\begin{equation}
    \left.\frac{\mathrm{d}\vec{v}}{\mathrm{d}t}\right|_{\rm Type\ I} = - \frac{\vec{v}}{\tau_a} - 2 \frac{(\vec{v} \cdot \vec{r})\vec{r}}{r^2 \tau_e},
    \label{eq:accel}
\end{equation}
where $\vec{r}$ and $\vec{v}$ are respectively the planet position and velocity.

The inner edge of the disk acts as a planet trap \citep{Masset2006} due to the sharp change in surface density.
We model a planet trap based on \citet{Pichierri2018} by dividing $\tau_a$ with a factor
\begin{equation}
\tau_{a,{\rm red}} =  \begin{cases}
     1, & a\geq r_{\rm ide}(1+h_{\rm ide}), \\
     5.5 \cos{\left(\frac{2\pi(r_{\rm ide}(1+h_{\rm ide}) - a)}{4h_{\rm ide}r_{\rm ide}}\right)} - 4.5, & |a -r_{\rm ide}| \leq r_{\rm ide}h_{\rm ide},\\
     -10, & a \leq r_{\rm ide}(1-h_{\rm ide}),
    \end{cases}\hspace{-0.5cm}
    \label{eq:tarev}
\end{equation}
which is activated only within a narrow distance from the exact inner disk edge position, $r_{\rm ie}$, measured as the scale height $h_{\rm ide}$ at $r_{\rm ide}$. 
The inner disk edge is set at $r_{\rm ide}=0.1\, {\rm AU}$ as suggested by \citet{Izidoro2017} and \citet{Brasser2018} to be common for solar type stars.
Both these effects were implemented via \texttt{Reboundx}, a complementary software which enables implementation of non-gravitational effects as forces to \texttt{Rebound} simulations \citep{Tamayo2020}.
The implementations used in this work have been made available as part of the \texttt{Reboundx} library\footnote{\href{https://github.com/dtamayo/reboundx}{https://github.com/dtamayo/reboundx}}.

Beyond a certain mass, a planet opens a partial gap in the disk, changing the nature of the migration  \citep{Lin1993,Ward1997,Crida2007,Baruteau2013,Kanagawa2015,Kanagawa2018}.
This regime is called type-II migration and the migration rate is no longer determined by Eq.~\eqref{eq:ta}.
Instead, the resulting torque onto the planet depends on the gap depth and the disk properties, in particular the turbulent viscosity parameterized by $\alpha$. 
\citet{Kanagawa2018} proposed a framework for gap opening planet migration where the migration rate from type-I can be amended to account for the gap by replacing the unperturbed gas surface density in Eq.~\eqref{eq:ta} by the surface density at the bottom of the gap.
The surface density inside the gap $\Sigma_{\rm{gap}}$ is related to the unperturbed gas surface density $\Sigma_{\rm{unp}}$ as
\begin{equation}
    \frac{\Sigma_{{\rm gap}}}{\Sigma_{{\rm unp}}}=\frac{1}{1+0.04K},
    \label{eq:gap-depth}
\end{equation}
where $K$ is given as
\begin{equation}
    K = \left(\frac{m_p}{M_*}\right)^2h^{-5}\alpha^{-1},
    \label{eq:K-eq}
\end{equation}
and $h$ is the gas scale height (Eq.~\ref{eq:h}) and $\alpha$ is the Shakura-Sunyaev turbulent viscosity parameter.

For values of $K$ larger than 20 \citep{Kanagawa2018}, the surface density is significantly affected and the migration is slowed by the gap.
From Eq.~\eqref{eq:K-eq}, the transition mass beyond which type-I migration is unrealistic is
 \begin{equation}
     \frac{m_{\rm{transition}}}{M_*}=8\times 10^{-5}\left(\frac{\alpha}{10^{-3}}\right)^{1/2}\left(\frac{h}{0.05}\right)^{5/2}.
     \label{eq:mig-transition}
 \end{equation}
To account for the transition to type-II migration, we replace in the expression of $\twave$ the surface density by $\Sigma_{\rm{gap}}$ given in Eq.~\eqref{eq:gap-depth} modifying both $\tau_e$ and $\tau_a$ as done by \cite{Kanagawa2020}.
The ratio $\tau_e/\tau_a$ remains unaffected by this modification.
While no semi-analytical expression exists for the eccentricity damping during type-II migration, this approach is supported by hydrodynamical simulations of showing a rapid damping of the eccentricity provided it remains small with respect to the disk aspect ratio \citep{Bitsch2010,Duffell2015}.

\subsection{Condition for resonant capture}
\label{sec:theory}

For each set of simulations ran, we analyse if the systems end up in a first-order MMR and which one.
At the end of the integration, all systems are either trapped in a MMR or became unstable.
We remove unstable systems by discarding systems where the final configuration is Hill unstable \citep{Marchal1982,Petit2018}.
Using the circular approximation \citep{Gladman1993}, Hill unstable systems verify
\begin{equation}
    a_2 - a_1 < 2\sqrt{3} R_{\rm H} = 2\sqrt{3} \frac{a_1+a_2}{2}\left(\frac{\ep}{3}\right)^{1/3},
    \label{eq:hillcrit}
\end{equation}
where 
\begin{equation}
{\ep = \frac{m_1+m_2}{M_*}}
\label{eq:epsilon}
\end{equation}
is the planet-to-star mass ratio.
The circular approximation is sufficient since systems out of MMRs circularize rapidly with respect to the migration timescale.

For a pair of planets exactly at the resonance $k+1$:$k$, we have
\begin{equation}
    kP_2-(k+1)P_1=0.
\end{equation}
By solving this equation for $k$, we determine the closest MMR by computing a continuous resonant index
\begin{equation}
    \kappa = \frac{1}{\frac{P_2}{P_1}-1}.
    \label{eq:k}
\end{equation}
Denoting by $[\kappa]$ the closest integer to $k$, we measure the distance to the resonance by computing \citep[\emph{e.g.}][]{Pichierri2019}
\begin{equation}
    \Delta = \frac{[\kappa]}{[\kappa]+1}\frac{P_2}{P_1} - 1.
    \label{eq:delta}
\end{equation} 
Here, $\Delta$ is positive for pairs of planets wide of the MMR.
We consider that a planet is trapped in a MMR if $0<\Delta<0.03$.
While such a criterion is not sufficient in general, we find that it gives a correct representation of the dynamical state while planets migrate through the disk. We integrate the system for a timescale at least 3 times larger than the time it takes to be trapped in a MMR.
If the planets were not trapped in MMR, the outer planet would cross the resonances leading to scattering events or collision with the inner one.
Furthermore, analysis of selected systems from our simulations show that the resonant angle librates for systems verifying the $\Delta$ criterion.

The theory of adiabatic invariants \citep{Henrard1982,Henrard1993} has shown that capture into resonance is in general a probabilistic event
\citep[see][and references therein for an up to date review]{Batygin2015}.
Yet, the particular structure of first-order MMR \citep{Henrard1983,Sessin1984} ensures that capture is guaranteed for convergent planets on close to circular orbits.
Indeed, at low eccentricity, it is possible to enter the resonance without crossing the separatrix.
The dynamics are considered adiabatic if a (quasi)-constant of motion of the non-dissipative problem  (\emph{i.e.} a parameter) evolves slowly with respect to the dynamical timescales of the rest of the system.
This is the case for the dynamics of two planets near a first-order MMR undergoing semi-major axis damping \cite{Henrard1983}.
However, eccentricity damping does not conserve the resonant variables which breaks the adiabatic assumption \citep{Goldreich2014}.

In the context of resonant motion, \cite{Batygin2015} proposed that adiabatic capture can take place if the resonance libration period is shorter than the migration time across the resonance
\begin{equation}
    \frac{ \tau_{\rm lib}  }{ \Delta t } = \frac{P_2}{\tilde{\tau}_a} \left(\frac{ M_* }{m_1+m_2}\right)^{4/3} \frac{1}{ 4k^{2/9}\left(\sqrt{3}|f_{\rm res}^{(1)}|\right)^{4/3}} \lesssim 1,
    \label{eq:adbc}
\end{equation}
where $\tilde{\tau}_a^{-1} = \tau_{a,2}^{-1}-\tau_{a,1}^{-1}$ is the relative migration rate and $f_{\rm res}^{(1)}$ is a function depending on the Laplace coefficients \citep{Laskar1995,Murray1999} that can be approximated at the resonance loci as $f_{\rm res}^{(1)}\approx -0.46 -0.802k$ \citep{Deck2013}.
While type-I migration is not an adiabatic process due to the eccentricity damping, 
the criterion \eqref{eq:adbc} provides a good analysis framework to understand the mechanism at play during resonance capture.
In this framework, the critical $\tau_{a,\mathrm{crit}}$ such that capture is possible scales as $\tau_{a,\mathrm{crit}}\propto \ep^{-4/3}$.
\cite{Ogihara2013} propose the same scaling based on numerical simulations.

In the context of type-I migration, the disk surface density acts as a proxy for the migration timescale.
As \citet{Batygin2015}, using Eq. \eqref{eq:adbc} and assuming the inner planet is trapped at the inner edge, we determine a critical surface density depending mainly on $k$, $\ep$ and $h_0$, below which capture into the $k$+1:$k$ MMR is possible
\begin{equation}
    \Sigma_0 \lesssim \Sigma^{(k+1:k)}_{\mathrm{0,crit}} = \ep^{1/3}M_* h_0^2\frac{a_2^{s+2\beta-2}}{r_0^{s+2\beta}} \frac{2k^{2/9}(\sqrt{3}|f_\mathrm{res}^{(1)}|)^{4/3}}{\pi(2.7+1.1s)(1+\zeta)^{-1}}.
    \label{eq:sigprop}
\end{equation}
where $r_0=1\, {\rm AU}$, $\zeta=m_1/m_2$.
The analytical theory predicts ${\Sigma_{\mathrm{crit}} \propto \ep^{1/3}}$, which should lead to a significant variation of the capture condition for a given surface density, if the planet mass is varied over several orders of magnitude.

When accounting for type-II migration, the simple functional form of Eq.~\eqref{eq:sigprop} can be preserved by replacing $\Sigma_0$ with 
\begin{equation}
    \Siggap = \frac{\Sigma_0}{1+0.04K}.
    \label{eq:siggap}
\end{equation}
In the chosen disk model, $K$ still depends weakly on the orbital radius since the disk is flared.
Neglecting this variation, we can treat the capture in MMR of gap opening planets in the framework of type-I migration, as if the planet is embedded in a reduced mass disk with a surface density given by $\Siggap$.

\section{Resonance capture for two-planet systems}\label{sec:results}

\subsection{Role of the surface density $\Sigma_0$ and the planet mass $\mpl$}\label{subs:s0_ep} 

\begin{figure}
    \centering
    \includegraphics[width=1.\linewidth]{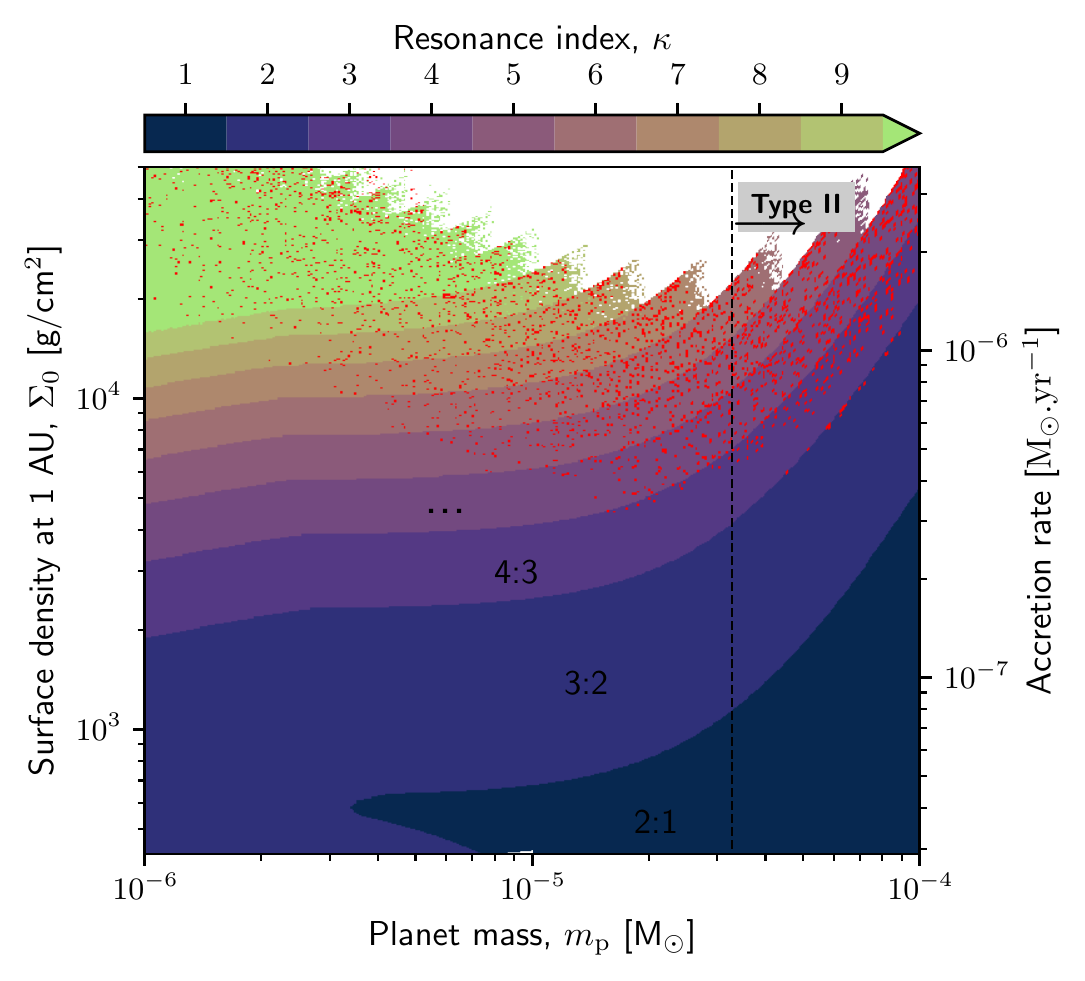}
    \caption{Final continuous resonance index $\kappa$ (Eq. \ref{eq:k}) of two equal mass planets as function of  the surface density $\Sigma_0$ at $1\, {\rm AU}$ and the planet mass $\mpl$. 
    The inner planet is fixed at the inner disk edge ($0.1\, {\rm AU}$) while the outer migrates from $0.2\, {\rm AU}$.
    Systems that do not satisfy $0<\Delta<0.03$ are considered not in resonance and are marked with red. 
    Hill unstable systems are removed and appear white. The dashed line indicates the transition mass (Eq.~\ref{eq:mig-transition}) to the type-II migration regime. 
    The accretion rate displayed on the right axis is computed from Eq.~\eqref{eq:Mdot} assuming a viscous parameter $\alpha = 10^{-2}$.}
    \label{fig:fiducial_tII}
\end{figure}
\begin{figure}
    \includegraphics[width=0.9\linewidth]{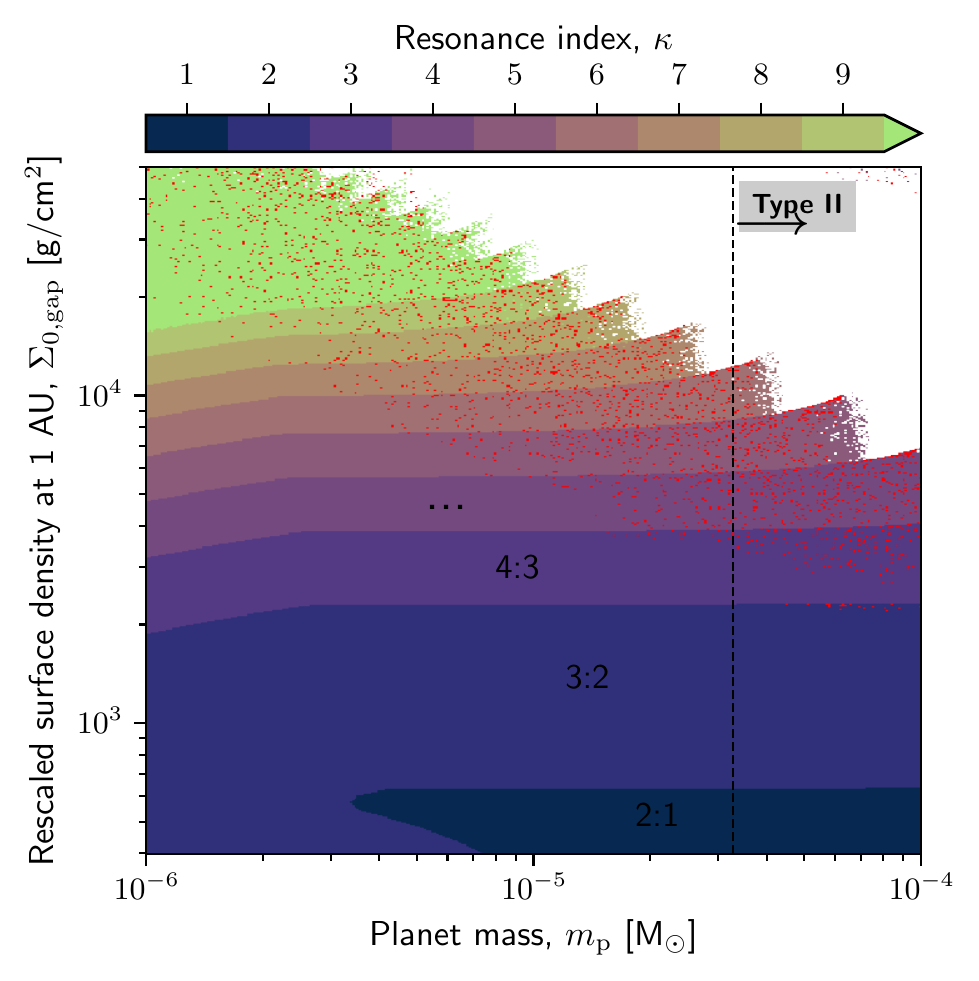}
    \caption{Final continuous resonance index $\kappa$ (Eq. \ref{eq:k}) of two equal mass planets as function of the rescaled surface density $\Siggap$ (Eq. \ref{eq:siggap}) and the planet mass $\mpl$. 
    Using the rescaled surface density allows to highlight the dynamically relevant quantity for the resonant capture.}
    \label{fig:fiducial}
\end{figure}
Based on the theoretical criterion \eqref{eq:sigprop}, we first focus on the impact on the capture in MMR from the surface density $\Sigma_0$ and the planet mass by varying the two over a wide range of values, even beyond the regime where type-I migration is valid. 
We vary the planet mass $\mpl = m_1=m_2$ over two orders of magnitude from $10^{-6}\, \Msun$ to $10^{-4}\ \Msun$.
For $\alpha=10^{-2}$ and $h_0=0.033$, Eq.~\eqref{eq:mig-transition} predicts that planets migrate in the type-I regime for masses smaller than $3.3\ 10^{-5}\ \Msun$.
The chosen value of $\alpha$ is consistent with our setting as the planets lie close to the inner edge of the disk where the magnetorotational instability is active and maintain the turbulence \citep[\emph{e.g.}][]{Johansen2006,Bai2011}. 
We vary the surface density at $1\, {\rm AU}$, $\Sigma_0$, from $4\times10^{2}\, \mathrm{g\, cm}^{-2}$ to $5\times10^{4}\, \mathrm{g\, cm}^{-2}$.
This range covers the typical surface densities of forming disks inferred from accretion rates \citep{Hartmann1998}, but is extended to higher values to probe the rapid migration regime where more compact resonances can be formed.
The differences in trends are more robust over a wider range and the whole picture becomes clearer.
In particular large $\Sigma_0$ values show the relationship between different resonances and probing large masses is necessary due to the weak exponent (1/3) of the analytical criterion.
Exploring the capture outcome beyond the regime compatible with observational constraints gives a consistency check in the comparison with the observations.
If the expected conditions during planet formation are compatible with the observations, whereas unphysical initial conditions lead to the formation of resonances that are not observed in exoplanet systems, we can conclude that the observed MMRs reinforce our constraints on the protoplanetary disk. Moreover, we benefit from the extended explored parameter space in our comparison between the analytical trends and the numerical simulations.
We run $170,000$ simulations exploring this 2D $(\mpl,\Sigma_0)$ parameter space.
As explained in \ref{subs:setup}, the planet masses are equal and the inner planet is fixed at the inner disk edge. 
In all the simulations, the aspect ratio at $1\, {\rm AU}$ is $h_0=0.033$ and we compute $K$ (eq. \ref{eq:K-eq}) at the initial position of the outer planet.

Figure \ref{fig:fiducial_tII} shows the final value of $\kappa$ (Eq. \ref{eq:k}) for the $170,000$ individual integrations as function of the surface density $\Sigma_0$ and the planet mass $\mpl$. 
We connect the surface density to the accretion rate of a viscous disk with Eq.~\eqref{eq:Mdot}, assuming $\alpha = 10^{-2}$ \citep{Hartmann1998,Hasegawa2017}.
Initial conditions where the outer planet encounters the Hill stability limit due to fast migration,  leading to unstable systems, are removed from the figure and appear white.
Red points correspond to systems out of resonance for the criterion \eqref{eq:delta}.
We first observe that for $\kappa \lesssim 10$, the transitions between the resonances are very clear and the resonant index increases with $\Sigma_0$.
Larger surface density leads to a faster migration bringing the planets to a smaller orbital separation (and thus larger resonant index) before capture becomes possible. 
For high resonant indexes ($\kappa \gtrsim 10$), the transitions are not as sharp due to resonance overlap.
These configurations only appear possible for unrealistic surface densities ($\Sigma_0\gtrsim 10^4\, \mathrm{g/cm}^{2}$) and so we do not analyse furthermore the results in this region. 

We note the transition mass $m_{\rm{transition}} =3.3\times 10^{-5}\Msun$ to the type-II regime (Eq.~\ref{eq:mig-transition}) with a vertical dashed line.
For a given surface density, planets with $\mpl\gtrsim m_{\rm{transition}}$ tend to be trapped in lower index resonances with respect to planets in a pure type-I migration regime due to the reduced migration rate.
For larger masses, we expect all pair of planets to be trapped in a 2:1 resonance.

A better insight onto the dynamics and MMR capture can be obtained by plotting the resonant index as a function of the rescaled surface density to account for the gap opening $\Siggap$.
Indeed, the migration rate is linear in $\Siggap$ and the planet mass $\mpl$ and this rescaling allows us to treat any system within the simpler type-I migration framework.
We plot on Figure~\ref{fig:fiducial} the same simulations as in Figure~\ref{fig:fiducial_tII} but as a function of $\Siggap$.

Inspired by the criterion \eqref{eq:sigprop}, we note $\Sigma^{(k+1:k)}_{\rm tr}(\mpl,h_0)$, the surface density below which a system is captured into the {$k$+1:$k$} resonance.
Following the method from \citet{Ogihara2013}, the transition surface densities $\Sigma^{(k+1:k)}_{\rm tr}$ are obtained by fitting for each $\mpl$ the condition $\kappa(\Sigma_0)<k+1/2$ as a probability function
\begin{equation}
    P(\kappa(\Siggap,\mpl)<k+1/2) = \left(1+\left(\frac{\Siggap}{\Sigma^{(k+1:k)}_{\rm tr}(\mpl)}\right)^{-\gamma_k}\right)^{-1}.
    \label{eq:fit-eq}
\end{equation}
As the transitions are sharp for low resonant indexes, the value of $\gamma_k$ is large and we do not report them. 

On Figure~\ref{fig:fiducial}, the transitions between the different resonances exhibit two different trends. 
In most of the investigated parameter range ($\mpl\gtrsim 3\times 10^{-6}\, \Msun$), for all $\kappa\leq10$, the values of $\Sigma^{(k+1:k)}_{\rm tr}$ show no dependency on $\mpl$ --and the planet-to-star mass ratio $\ep$-- and occur at fixed surface density values for all $\kappa$.
The transition surface density is constant across two order of magnitude in $\mpl$.
According to the capture criterion \eqref{eq:adbc}, we should have $\Sigma^{(k+1:k)}_{\rm tr}\propto \ep^{1/3}$ and thus observe a significant change of transition surface density with the planet mass.
For planets in the regime of masses that is relevant for type-I migration, our simulations show the limits of the adiabatic framework.

For $\mpl\lesssim 3\times 10^{-6}\, \Msun$, the transitions show a weak $\mpl$ dependency.
Fitting $\Sigma^{(k+1:k)}_{\rm tr}$ as a power law of $\mpl$ for values smaller than $3\times 10^{-6}\, \Msun$ and $2\leq \kappa\leq10$, we get $\Sigma^{(k+1:k)}_{\rm tr} \propto \ep^{0.19\pm0.01}$, which is still smaller than the $1/3$ analytical value. 
It remains possible that at lower masses, the transition power exponent could match the analytical criterion.
Yet, this regime is not relevant for planet formation as planets grow to larger masses before experiencing significant migration \citep{Johansen2019}.

For low mass planets ($\mpl\lesssim 3\times10^{-6}\, \Msun$), the transition between the 2:1 and the 3:2 MMR display a different pattern than higher index transitions as there are no systems in the 2:1 MMR for these masses. 
We interpret this feature as a sign of the overstability of the 2:1 resonant center in this regime.
The linear stability analysis of the resonance fixed point \citep{Deck2015} has shown that the resonant variable can spiral out and eventually escape the resonance.
The stability of the fixed point is mainly governed by the mass ratio of the planets (a smaller inner planet can lead to overstability while a more massive inner planet usually stabilizes the resonance), the specific resonance (the 2:1 MMR is more sensitive than higher indexes) and the ratio of eccentricity dissipation between the two planets (which in the case of type-I migration is roughly proportional to the planet mass ratio).
\citet{Xu2018} have shown that the more realistic migration and eccentricity damping rates from \citet{Cresswell2008}  can allow overstable configurations to remain trapped in the MMR but with a finite libration amplitude.
We note that the shape of the transition we observe on Figure \ref{fig:fiducial} is similar to the shape of the overstable region in the Figure 3 of \citet{Xu2018}.
In the case of resonance escape, the planets resume migration towards inner MMRs until a stable MMR is encountered \citep{Deck2015}. 
In this case the planets escape 2:1 and stabilize in 3:2.
In Appendix \ref{app:overstability}, we analyse further the role of the overstability of the resonance by considering the case of two planets with unequal mass, as well as the case of two migrating planets.
As predicted by \cite{Deck2015}, our simulations show that the overstability occurs for all resonances when the outer planet is more massive than the inner one provided that the planet-to-star mass ratio is smaller than a critical value.

\subsection{Influence of the disk aspect ratio $h_0$}
\label{subs:s0_h0}
\begin{figure}
        \centering
        \includegraphics[width = \linewidth]{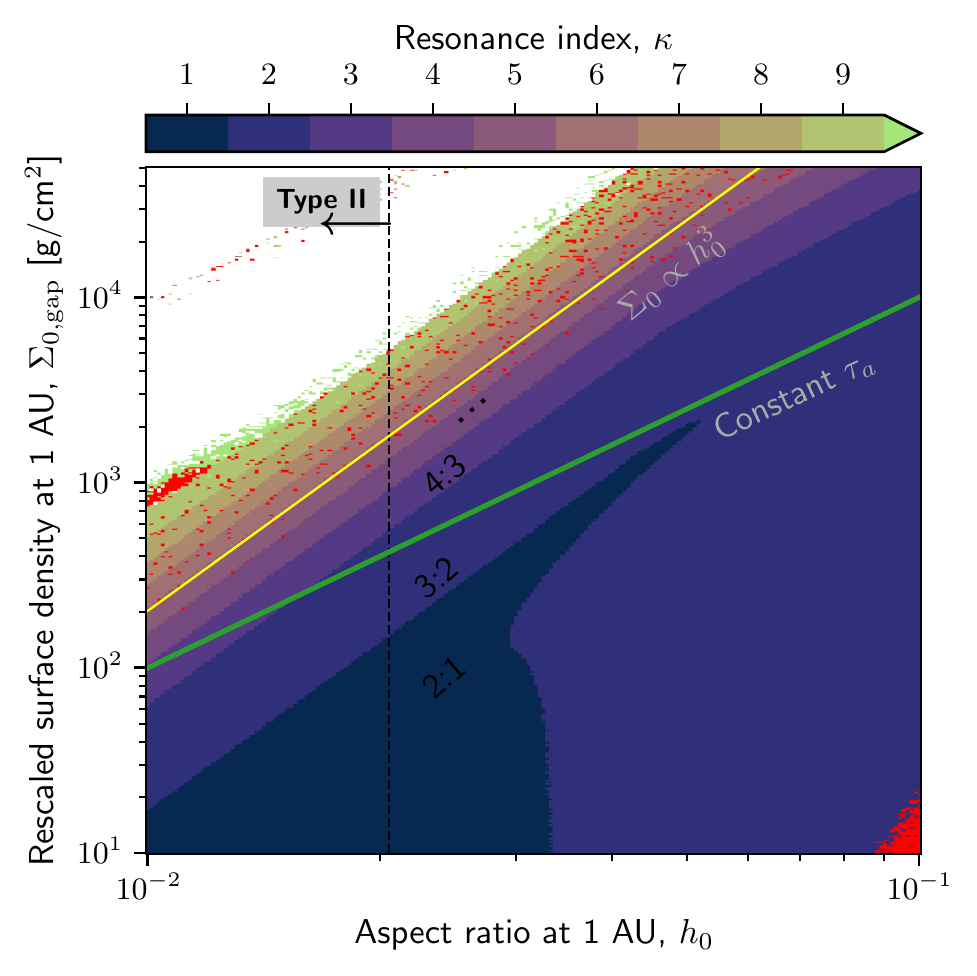}
        \caption{Continuous resonance index $\kappa$ for a pair of $10^{-5}\, \Msun$ equal mass planets as function of the rescaled surface density $\Siggap$ (Eq. \ref{eq:siggap}) and the disk aspect ratio $h_0$.
        The green line verifies $\Sigma_0 \propto h^2$, corresponding to a constant migration timescale.
        The yellow line is an example of the fitted $\Sigma^{(k+1:k)}_{\rm tr} \propto h_0^3$. 
        The black dashed line indicates the gap opening threshold, $h_0 = 0.02$.
        See Figure~\ref{fig:fiducial_tII} for a complete description.}
    \label{fig:h0vs0}
\end{figure}
\begin{figure*}
   \centering
   \includegraphics[width = \linewidth]{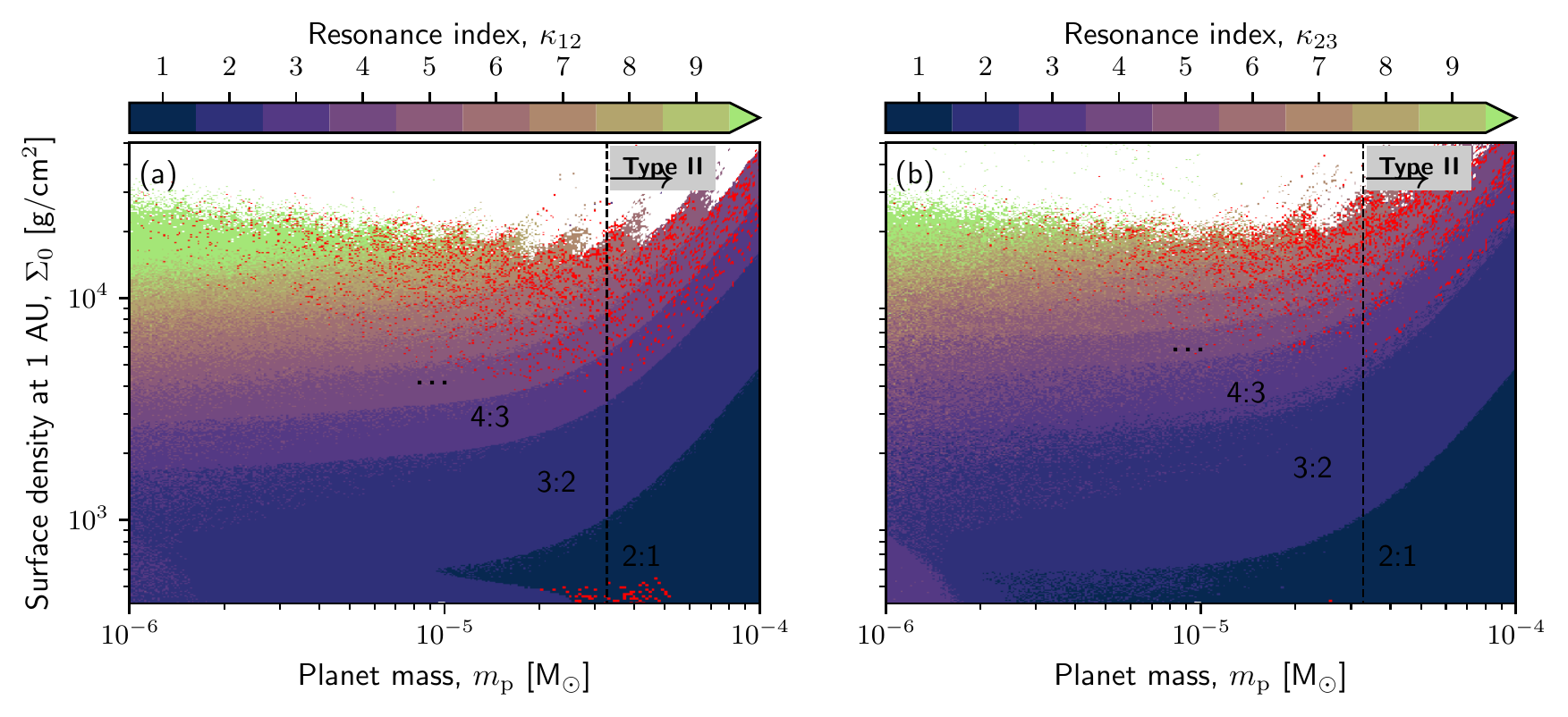}
   \caption{Continuous resonant indexes $\kappa_{12}$ (a) and $\kappa_{23}$ (b) of the inner and outer pair of planets as a function of $\mpl$ and $\Sigma_0$ for three equal mass planet systems. 
   The innermost planet is initialized at the inner disk edge while the other two migrate respectively from $0.2$ and $0.3\, {\rm AU}$. 
   The initial conditions are otherwise identical to the first run presented in Figure \ref{fig:fiducial}. The black dashed line indicates the gap opening threshold at $3.3\times 10^{-5}\Msun$.}
   \label{fig:3p}
\end{figure*}
For planet undergoing type-I migration, we showed that the planetary mass has a weak impact on MMR capture. 
We now keep the planetary mass constant at $\mpl = m_1=m_2 = 10^{-5}\, \Msun$ (below the type-II transition mass for $h_0=0.033$) and instead investigate the role of the disk aspect ratio, $h_0$, onto the final resonant configuration.
From Eq. \eqref{eq:ta}, we have that $\tau_a\propto h_0^2/\Siggap$ so systems with equal migration timescales lie on curves verifying $\Siggap\propto h_0^2$.
According to the criterion \eqref{eq:adbc}, we expect the resonance transitions to have a similar dependency.
When we vary $h_0$, we also change the relationship between $\tau_e$ and $\tau_a$ since $\tau_e/\tau_a \propto h_0^2$. As a result, we can test if $\tau_e$ impacts the MMR capture.
The disk aspect ratio is much more constrained than the other considered parameters.
Understanding its influence onto the MMR capture is thus more relevant from a theoretical perspective.
As we have seen in our previous setup, the transition to the type-II regime tends to hide the dynamical features such as the power-law trend in the transition surface densities.
We thus decide to present the results of this section only as a function of the rescaled surface density $\Siggap$.
The transition to the type-II regime can also be expressed as a critical aspect ratio.
For a planet of mass $\mpl =10^{-5}\, \Msun$ at $0.2\, {\rm AU}$, the planet opens a gap if the aspect ratio $h_0 \lesssim h_{0,\rm{transition}}= 0.02$.

We vary the rescaled surface density $\Siggap$ from $10\, \mathrm{g\, cm}^{-2}$ to $5\times 10^{4}\, \mathrm{g\, cm}^{-2}$ and $h_0$ from $0.01$ to $0.1$, simply spanning an order of magnitude around the typical value taken as realistic in the literature. 
We integrate $80,000$ equal mass two-planet systems and present the results in a 2D grid, where each couple corresponds to a combination of $h_0$ and $\Siggap$. 
Again, the inner planet is initialized at the inner disk edge and the planets have the same mass.

Figure \ref{fig:h0vs0} shows the final value of $\kappa$ as function of $\Siggap$ and $h_0$. 
Similarly, the white points correspond to unstable systems and the red dots are systems considered out of resonance. 
The green line verify $\Siggap \propto h_0^2$ to help visualize configurations with an equal migration rate.
We see similarities with Figure \ref{fig:fiducial} in the behaviour of $\Siggap$ and the sharpness of the transitions between the resonances. 
Larger surface density $\Siggap$ or a thinner disk (smaller $h_0$) leads to faster migration and so capture into higher $k$ MMRs.
The transitions are sharp for indexes $\kappa\lesssim 10$.
For the thicker disks ($h_0>0.033$), we observe a break in the power law trend as no systems end up in the 2:1 MMR due to the overstability of the fixed point.

We fit the transition between the resonances and we get that $\Sigma^{(k+1:k)}_{\rm tr} \propto h_0^{3.02\pm0.01}$, instead of the scaling $\propto h_0^2$ expected from the analytical modeling.
The exponent $3.02\pm0.01$ suggests that the eccentricity damping also plays a role in the capture in MMR.
This result is surprising since \citet{Ogihara2013} found no significant trend when looking at the impact of $\tau_e$ on the transition between capture in different MMRs.
A possible explanation may be that \citet{Ogihara2013} did not consider the case where both planet's eccentricity damping vary at the same time.

\section{Three-planet resonant chains}\label{sec:3p}

Resonant chains contain often more than two planets.
It is thus natural to test whether the results found in previous subsections hold for more complex systems.
We use a three planet coplanar system with the same conditions as in the run presented in Subsection \ref{subs:s0_ep}, to see if the qualitative behaviour and relationship between $\Sigma_0$ and $\mpl$ change.
In this run, the innermost planet is fixed at the inner disk edge at $0.1\, {\rm AU}$, the middle one migrates inward from $0.2\, {\rm AU}$, and the outermost one from $0.3\, {\rm AU}$.
The capture is thus sequential, the first pair of planets is captured before the outer planet joins the chain, which can eventually affect the inner pair. 
The initial parameters $\Sigma_0$ and $\mpl$ span the same 2D grid.

We plot on Figure \ref{fig:3p} the resonant indexes $\kappa_{12}$ and $\kappa_{23}$ for the respective planet pair. 
The overall qualitative behaviour is preserved as we do not observe a dependency on $\mpl$ for $\Sigma_{\rm tr}^{(k+1:k)}$ before the transition to the type-II regime.
However, the transitions between the resonances are noisier, particularly for the outer pair.
Since the outer planet is captured while the inner pair of planets is already in resonance, the middle planet is no longer on a circular orbit which makes the capture of the outer planet no longer deterministic \citep{Henrard1982,Batygin2015}. 
We can note that the resonant index is more stochastic for smaller planetary masses, probably due to the fact that the resonance islands are smaller.

The inner pair can also be disrupted by the subsequent capture of the outer pair.
We note that the surface density transitions $\Sigma_{\rm tr}^{(k+1:k)}$ for the inner pair are on average 22\% lower than the same transition in the two-planet case (see the next section \ref{sec:s0_k}, Figure \ref{fig:all_sig(k)} and Appendix \ref{app:comp2v3}).
The difference between the resonant indexes of the inner pair and the two-planet result from section \ref{subs:s0_ep}, is displayed on Fig.~\ref{fig:k12-k2p}.
We find that, within the considered parameter space, 36\% of the inner pairs are trapped in the next index resonance compared to the two-planet case.
By studying the inner pair resonant indexes before the capture of the outer planet, we note that the inner pair behaves similarly to the two-planet case.
However, the originally formed resonance can be disrupted by the arrival of the outer planet for systems close to the surface density transition.
In this case, the migration resumes until the inner pair is captured into the next resonance.
The picture is slightly different for the outer pairs. We plot on Fig.~\ref{fig:k23-k2p} the difference between the outer pair resonant indexes and the two-planet system resonant indexes.
We find that the transitions occur almost at the same surface densities as in the two-planet case but with a significant scattering around the transition, with 2\% of the pairs trapped in a tighter resonance and 26\% in a wider one with respect to the two-planet case.

\begin{figure}
   \centering
   \includegraphics[width = \linewidth]{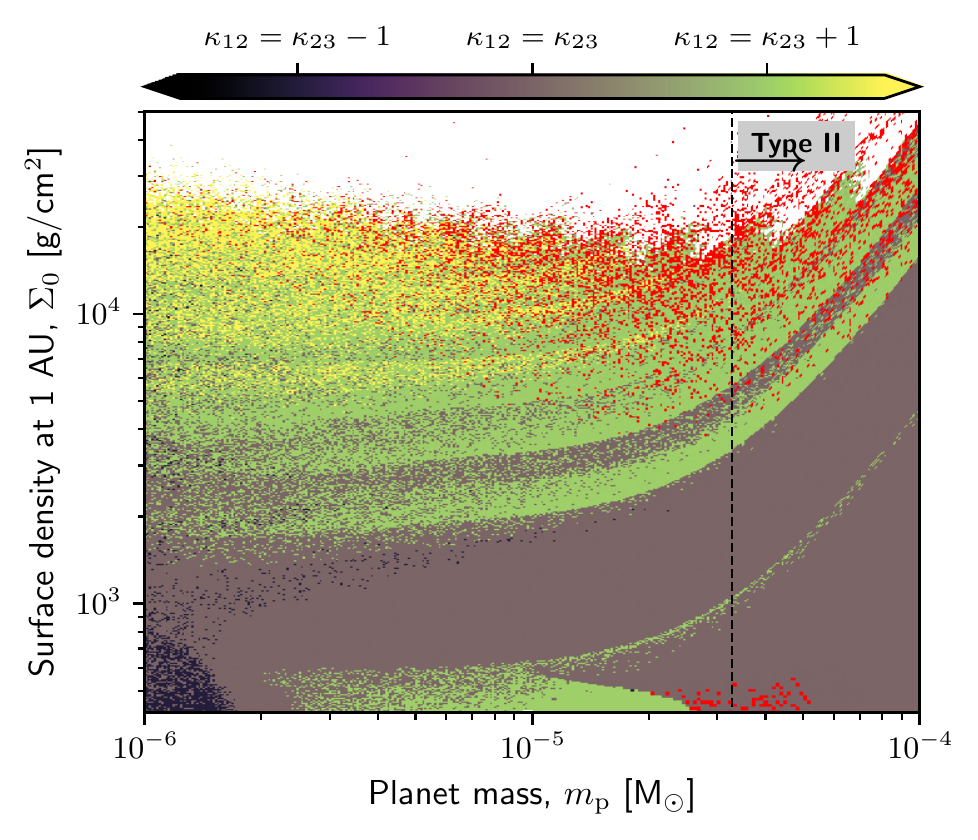}
   \caption{Difference $\kappa_{12} - \kappa_{23}$ of the inner and outer pair continuous resonant indexes, as a function of the planet mass $\mpl$ and the surface density $\Sigma_0$. 
   As in previous figures, the red dots correspond to systems far from resonance, and the black dashed line correspond to the gap opening threshold ($3.3\times10^{-5}\Msun$).}
   \label{fig:3pdiff}
\end{figure}

We also quantify the intra-system uniformity. 
We plot on Fig. \ref{fig:3pdiff}, the difference of resonant index $\kappa_{12} - \kappa_{23}$ between the inner and the outer pair of each system.
For the considered parameter space and among all systems where both pairs of planets are in resonance, we find that only 39\% form a uniform chain with $\kappa_{12}=\kappa_{23}$.
For non uniform chains, we find that $\kappa_{12} = \kappa_{23}+1$ in 57\% of the cases and $\kappa_{12} = \kappa_{23}+2$ in 3.3\% of the cases. 
Systems where the inner pair is wider than the outer one amounts for less than 1\% of the chains.
However, this general picture is slightly misleading when one focuses on the low index resonances.
Indeed, we find that whenever the inner pair is in a 2:1 resonance, all the systems form a Laplace 4:2:1 chain.
Similarly, 98\% of the systems where the inner pair is in a 3:2 resonance are 9:6:4 resonant chains.
On the other hand, only 16\% of chains where the inner pair is trapped into a higher index ($\kappa_{12}\geq3$) resonance are uniform chains.

Our findings show that the two-planet results from the previous subsections can be generalized to more complex systems.
In particular, we interpret uniform chains as a consequence of the disk properties being the dominant parameters leading to the capture into a specific resonance.
We expect that systems with four planets or more would also show the same qualitative behaviour.
However, resonant systems with many planets can become unstable due to excitation from secondary resonances, complicating the picture.
The mass threshold for this type of instability decreases with increasing $k$ and with increasing number of planets \citep{Pichierri2020}.

\section{Transition surface density $\Sigma_{\mathrm{tr}}$ as function of the resonance index $k$}\label{sec:s0_k}
\begin{figure}
    \centering
    \includegraphics[width=\linewidth]{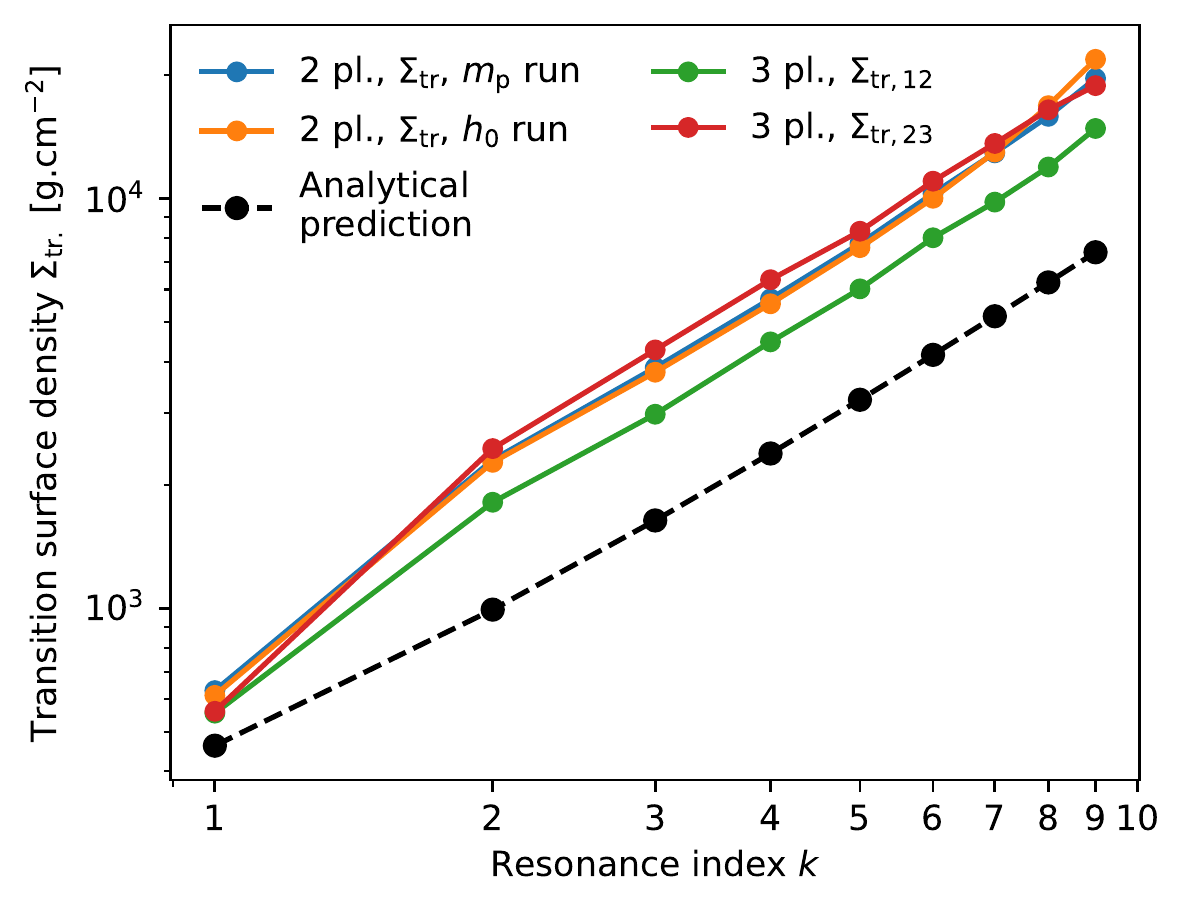}
    \caption{Transition surface densities $\Sigma^{(k+1:k)}_{\rm tr}$ as a function of the resonant index $k$ for the different set-ups plotted in Figure \ref{fig:fiducial}, Figure \ref{fig:h0vs0} and the two grids in Figure \ref{fig:3p}.
    We estimate the transition surface densities on the planet mass $\mpl$ independent regime for the first set-up and the three-planet case.
    For the varying $h_0$ data, we model the $h_0$ dependency as $\Sigma_0\propto h_0^3$ and record the value for $h_0=0.033$.
    The black points correspond to the analytical estimate Eq.~\eqref{eq:sigprop} for equal mass planets, $\mpl=m_1=m_2=10^{-5}\ \Msun$ and $h_0=0.033$. }
    \label{fig:all_sig(k)}
\end{figure}
We combine the results from our different set-ups to study the trend in the resonant index for the transition surface densities.
Eq.~\eqref{eq:sigprop} predicts that $\Sigma^{(k+1:k)}_{\rm tr}$ scales asymptotically as $ k^{1.55}$.
However, for $2\leq k \leq10$, a power-law fit of Eq.~\eqref{eq:sigprop} leads to a slope of $1.34$.
We exclude the 2:1 resonance from the trend analysis due to its unique dynamics \citep{Beauge1994}.
We fit $\Sigma^{(k+1:k)}_{\rm tr}$ as described in Section \ref{subs:s0_ep} for the results presented above. 
From the setup displayed in Figure~\ref{fig:fiducial} and the three-planet systems, we use the constant transition regime for $\mpl\gtrsim10^{-5}\, \Msun$.
As seen in Section \ref{subs:s0_h0}, the transition surface density is not constant with $h_0$ but we have $\Sigma^{(k+1:k)}_{\rm tr}\propto h_0^3$.
We therefore model the aspect ratio dependency from the Section \ref{subs:s0_h0} results by taking the average value of $\Sigma^{(k+1:k)}_{\rm tr}(h_0/0.033)^{-3}$ such that the reported value is comparable to the other setups.
The chosen planet mass and disk aspect ratio places us in the type-I regime where we can neglect the effect of the gap opening.

We plot in Figure~\ref{fig:all_sig(k)}, the transition surface densities as a function of $k$ from our simulations as well as the analytical prediction Eq. \eqref{eq:sigprop} for $\ep=2\times 10^{-5}$ and $h_0 =0.033$.
The analytical prediction is smaller by a factor 2 to 3 than the numerical fits, but the actual factor is mostly determined by the choice of aspect ratio and $\mpl$ since the scaling found numerically differs from the analytical one. 
We observe that all results follow a similar power-law trend for $k\geq2$ with a slope of $1.42\pm0.04$, which is consistent with the analytical prediction.
This agreement shows that the dynamics of the resonant motion are likely well approximated by the first-order model used in \citep{Batygin2015}.
On the other hand, the discrepancy in the mass and aspect ratio trends suggests that the migration timescale is not the only parameter governing the resonance capture but that the eccentricity damping should play a significant role.
Similarly, the analytical surface transition being smaller than the numerical fit can be interpreted by noting that the actual capture mechanism is more robust than the adiabatic condition predicts.

Moreover, we see on Figure~\ref{fig:all_sig(k)} that the transition surface density $\Sigma_{\mathrm{tr},12}$ for the inner pair of planets in the three planet case is lower than in the two-planet case.
As explained in Section \ref{sec:3p}, the outer planet compacts the inner pair when it enters in resonance with the middle planet.

\section{Comparison to observed chains}\label{sec:discussion}
\begin{figure}
    \centering
    \includegraphics[width = \linewidth]{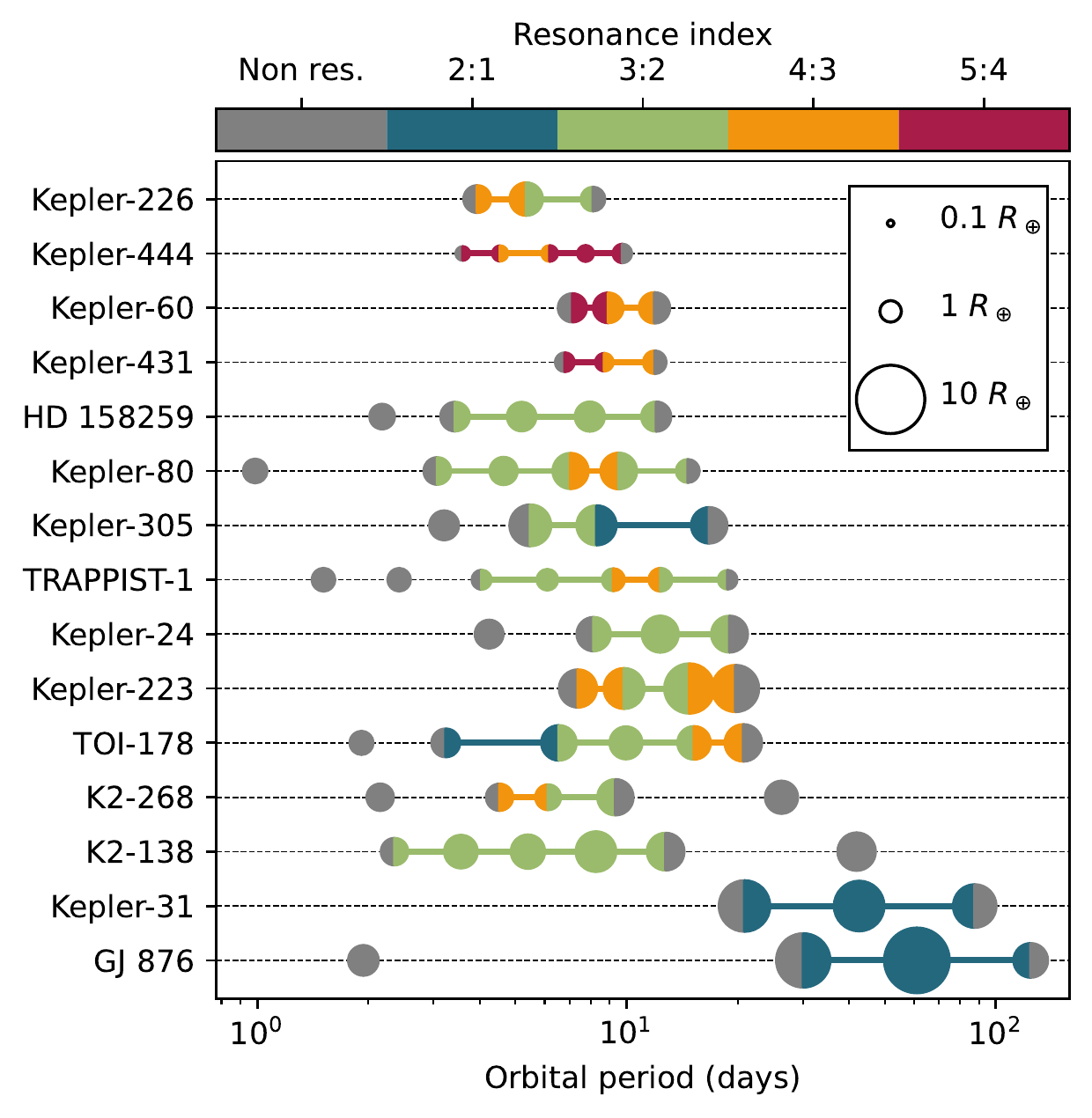}
    \caption{The distribution of first-order MMRs among exoplanet-systems in 3+ planet resonant chains as a function of their orbital period. 
    The color indicates the first-order MMR in which resonant planet pairs are.
    See Section \ref{sec:discussion} for the selection process.
    These resonant chains are consistent with a formation by type-I migration. As discussed in Section \ref{sec:3p}, low index (2:1 and 3:2) MMRs tend to belong in uniform resonant chains while at higher indexes, the inner pair is usually in a higher index MMR with respect to the outer one.}
    \label{fig:exoplanet}
\end{figure}
Our results show that for planets below the type-II mass transition, capture in resonance at the inner edge of a disk is mainly ruled by the disk parameters rather than the planet properties (although in the case of an inner less massive planet, fix point overstability can play a role as seen in Appendix \ref{app:overstability}).
As seen in Figure~\ref{fig:fiducial}, for type-I migrating equal mass planets, we see close to no dependency in the planet mass.
In the adiabatic framework, the MMR capture is possible if the migration time across the resonance is longer than the resonance libration period \citep{Batygin2015}.
We showed that this model provides a qualitative explanation for the transition between successive resonances.
However, we see that the eccentricity damping likely plays a significant role as capture is also possible for systems where the libration timescale is longer than the resonance crossing time.
Beyond two-planet systems, we see that the capture properties are not strongly affected in longer chains.
For an inner pair of planets with a low resonance index (2:1 or 3:2 MMR), the chain is most likely uniform, \emph{i.e.} with the outer planet pair having a similar index.
For higher index, the inner pair is usually tighter than the outer pair.
We can use these results to link the observed exoplanets to the conditions of formation of resonant chains. 

While our simulations are performed with a planet trapped at the inner disk edge at $0.1\, {\rm AU}$ around a solar mass star, it is quite straightforward to rescale the problem to generalize our results.
Changing the stellar mass amounts to changing the mass units of our simulations and the surface density and planet masses should be rescaled accordingly.
Changing the mass units while keeping the same gravitational constant and time units, requires the modification of the length unit.
Capture beyond the inner disk edge at an arbitrary distance can also be predicted from our simulations by using the surface density at the capture location instead of its value at $0.1\, {\rm AU}$. 

Planet migration is a well understood process that must take place in the protoplanetary disk from its formation  to its photoevaporation.
Yet, during the early growth phases, planets mostly grow while barely moving as the growth timescale is shorter than the migration one \citep{Johansen2019}.
The growth from lunar mass embryos to super-Earths is governed by the pebble mass flux in the disks inner region, low flux means slow growth thus little migration during the disk lifetime.
In particular, \cite{Lambrechts2019} showed that super-Earth embryos start to migrate only after half a million year.
The typical surface densities at $1\, {\rm AU}$ during the disk lifetime can be derived from the disk accretion rate.
If we assume a viscous spreading parameter $\alpha=10^{-2}$, a typical Class II disk surface density at $1\, {\rm AU}$ ranges from $ \simeq2000\, \mathrm{g\, cm^{-2}}$ to $\simeq10\, \mathrm{g\, cm^{-2}}$, when the accretion rate drops from a few $10^{-7}\, \mathrm{M_\odot\, yr^{-1}}$ to $10^{-9}\, \mathrm{M_\odot\, yr^{-1}}$ over the life-time of the protoplanetary disk.
From Figure~\ref{fig:fiducial_tII}, we see that type-I migration in this range of surface densities leads to trapping into 3:2 or 2:1 MMRs for most systems. 
In particular, 3:2 MMR are formed for $\dot{M}$ between $3\times10^{-8}\, \mathrm{M_\odot\, yr^{-1}}$ to $2\times 10^{-7}\, \mathrm{M_\odot\, yr^{-1}}$, which corresponds roughly to the first $1\, {\rm Myr}$ of the disk lifetime \citep{Testi2022}.
Within the inner parts of the disk that we consider here, there is more than enough time to grow protoplanets until they start migrating within that time-frame.
Our simulations provide a loose evidence that super-Earth formation likely happens early in the evolution of the protoplanetary disk.
One should note that the surface density of protoplanetary disks are not very well constrained, due to the weak constraints on the viscous parameter $\alpha$ or the role of magnetic winds \citep{Tabone2022} that can affect the profile of the disk.
In particular, slightly higher surface densities are not ruled out, which could also explain the observed higher indexes resonances as a result of planet formation at later evolutionary stages.

Next, we compare our results to the observed multiple exoplanet resonant chains using the NASA exoplanet archive\footnote{\href{https://exoplanetarchive.ipac.caltech.edu/}{https://exoplanetarchive.ipac.caltech.edu/}}.
Following the same approach as \citet{Pichierri2019}, we select all systems that contain a pair of planets with $0<\Delta<0.03$.
The selection condition only enforces that the planet pairs are slightly wide of commensurability and not that they are actually in resonance.
However, detailed dynamical analysis of some of these systems, in particular the longer chains, have shown that they are still in resonance (\emph{e.g.} Trappist-1, \citealp{Agol2021} or TOI-178, \citealp{Leleu2021}).
Moreover, tidal effects during the system lifetime can lead resonant systems to settle slightly wide to the resonance configuration where they were originally captured \citep{Delisle2014,Millholland2019}.
Given the fragility of resonant chains \citep{Izidoro2021,Raymond2021}, we can expect these chains to be close to their original state. 
We find 158 pairs close to a first-order MMR within 127 systems.

The majority of the observed exoplanet pairs close to MMR are isolated (\emph{i.e.} not part of a longer chain).
Among them, we find 52 pairs close to the 2:1 MMR, 67 pairs close to the 3:2 MMR, and 39 close to a higher index resonance.
This distribution of MMR is consistent with the capture of planets undergoing migration in a typical protoplanetary disk as the surface density needed for the capture in a $k\leq2$ MMR is comparable to the typical surface density in a protoplanetary disk.  
However, it is harder to connect a specific pair to the disk properties at the time of the formation.
Instead of looking at individual resonant pairs, we now focus on three and more planet chains, as we can study how the trends in an individual system compare to the results from Section \ref{sec:3p}. 
As a result, we only keep the systems containing two consecutive resonant pairs.

The 15 selected systems are plotted in Figure~\ref{fig:exoplanet} as a function of their orbital period.
The size of the planet corresponds to their radius when available or we define the planet radius as $R_p = R_\oplus (M_p/\Mearth)^{1/3}$ for illustrative purposes.
If a pair of planets is resonant, we link the two planets with a line segment of color according to the resonance index.
We leave the planet grey if it is out of resonance.

We can divide the selected systems into three broad categories.
There is a system subset composed of low index (3:2 and 2:1) uniform resonant chains (HD 158259, Kepler-24, K2-138, Kepler-31, GJ 876).
We observe that the 4:2:1 chains are composed of massive planets.
Kepler-31 b, c and d have a radius of respectively $5.5\, \mathrm{R_\oplus}$, $5.3\, \mathrm{R_\oplus}$ and $3.9\, \mathrm{R_\oplus}$ \citep{Fabrycky2012}.
GJ 876 c, d and e have reported minimum masses of respectively $227\, \Mearth$, $723\, \Mearth$ and $14\, \Mearth$ \citep{Rivera2010}.
In both cases, the two innermost planets of the chains have masses or radii comparable to Saturn or Jupiter.
These planets likely opened gap in their protoplanetary disk which slows their migration and allow the capture in the 2:1 resonance \citep{Kanagawa2020} as shown on Figure \ref{fig:fiducial_tII}.

We also see three-planet chains with a higher index inner pair \emph{i.e.} chains of the form $k$+2:$k$+1:$k$ (Kepler-226, Kepler-60, Kepler-431, Kepler-305, K2-268).
In these systems, the inner pair can initially be in the same resonance as the outer pair before being pushed into the higher index MMR during the capture of the outer planet.
These two categories are consistent with the simulations presented in Section \ref{sec:3p}.
We see that uniform chains tend to have low resonant indexes whereas higher index resonances belong to $k$+2:$k$+1:$k$ chains.

Finally, the latter category is composed of the systems out of the previous two patterns (Kepler-444, Kepler-80, Trappist-1, Kepler-223, TOI 178).
These systems have a more complex architecture with longer chains.
As resonant planets have a non-zero eccentricity, the capture of a new planet in an already existing chain can become probabilistic \citep{Batygin2015} and create non-uniform chains.
Yet, the existence of these systems is not incompatible with our results.
Indeed, in most of these systems, we only observe a single pair that differs from the other resonances.
Moreover, detailed studies on these particular systems have shown that these systems are consistent with a formation under type-I migration (\emph{e.g.} Kepler-223, \citealp{Huhn2021}).
Post-disk phase dynamics can also modify the architecture of the systems.
Tidal effects are particularly important for such systems because of their short inner orbit and the resonance boosting the planet eccentricity, fueling the dissipation and migration of the inner pair \citep{Delisle2014}.
As an example, Trappist-1 was most likely  originally a seven planet resonant chain composed of only 3:2 MMR just as expected from our simulations, except for the 4:3 pair between f and g.
The two innermost planets experienced orbital decay such that they no longer are in two-body resonance, but a three-body resonance remains as it is not disrupted by tidal decay \citep{Huang2022}.
A similar mechanism may have taken place in TOI-178 \citep{Leleu2021}.

\section{Conclusions}\label{sec:concl}

Our results provide a more comprehensive view onto the capture into resonance in the context of planet formation.
We generalize the results from previous numerical studies \citep{Quillen2006,Ketchum2011,Ogihara2013,Xu2018} and confirm some of the trends already known, while also putting into light new behaviours thanks to the widely probed parameter space.
We staged our experiments in a theoretical context, laid out by the analytical works of \cite{Henrard1983}, \cite{Batygin2015} and \cite{Deck2015}.
The adiabatic resonant capture is possible assuming that the migration timescale through the resonance is longer than the resonant libration period.
We confirm that this theoretical framework provides a good description of the mechanism leading to resonant capture, as our results are qualitatively similar to the predictions.
However, we show that the capture for planets undergoing type-I migration remains a non-adiabatic mechanism such that the quantitative trends derived from the analytical criterion of \cite{Batygin2015} are not verified.
In particular the eccentricity damping favours the MMR capture and plays a significant role, as highlighted by the results of Section \ref{subs:s0_h0}.
This, as well as the absence of dependency of the transition surface density in the planetary masses (in the restricted type-I framework) is in contradiction with previous results such as the one obtained by \cite{Ogihara2013}.
We conjecture that the use of a more realistic migration prescription \citep{Cresswell2008} is responsible for this discrepancy as already hinted by \citet{Xu2018}. 
The critical importance of the non-adiabatic mechanisms motivates further theoretical works.

We compare our simulations to the architecture of exoplanet resonant systems.
We find that most of the observed resonant pairs have a low resonant index ($k\leq2$) which is consistent with the capture criteria for planets undergoing migration in a typical protoplanetary disk.
This is a strong evidence that planetary systems are shaped by the migration. 
Furthermore, we can look at the structure of longer chains as their architecture can be confronted to our simulations without knowing the precise conditions during formation. 
Indeed, resonances within the same system formed under the same conditions.
We find that planets composed of low index resonances are typically in uniform chains while higher-index resonances are more often in chains of the form $k$+2:$k$+1:$k$, in agreement with our numerical findings.

Furthermore, as this work can be used to link current architecture and properties of a system to its formation conditions, it is of interest to further build on this study with more detailed models of migration, disk or the inner disk edge. 
That would take us closer to fully understanding MMR as an important part of planetary evolution. 

\begin{acknowledgements}
    This research has made use of the NASA Exoplanet Archive, which is operated by the California Institute of Technology, under contract with the National Aeronautics and Space Administration under the Exoplanet Exploration Program.
	This research was made possible by the open-source projects \texttt{iPython} \citep{Perez2007}, \texttt{Jupyter} \citep{Kluyver2016}, \texttt{matplotlib} \citep{Hunter2007}, \texttt{numpy} \citep{vanderWalt2011}, \texttt{scipy} \citep{Virtanen2020}, \texttt{pandas} \citep{WesMcKinney2010}, \texttt{Rebound(x)} \citep{Rein2012,Tamayo2020}, and \texttt{xarray} \citep{Hoyer2017}.
	A.J. and A.C.P. acknowledges funding from the European Research Foundation (ERC Consolidator Grant 724687-PLANETESYS), the Knut and Alice Wallenberg Foundation (Wallenberg Scholar Grant 2019.0442), the Swedish Research Council (Project Grant 2018-04867), the Danish National Research Foundation (DNRF Chair Grant DNRF159) and the Göran Gustafsson Foundation.
\end{acknowledgements}

\bibliographystyle{aa} %
\bibliography{Article2021}

\begin{appendix}

\section{Overstability of the MMR in the case of a more massive outer planet}\label{app:overstability}

Planets captured in resonances reach an equilibrium point as eccentricity damping balances the convergent migration \citep{Papaloizou2005}.
However, this equilibrium point can be linearly unstable such that the resonant variable can spiral out of the resonance \citep{Papaloizou2010,Xu2018} by a mechanism called overstability \citep{Deck2015}. 
After the breaking of the resonance, the outer planet continues migrating towards an inner resonance. \citet{Deck2015} found that the resonant fixed point is overstable if the total planet mass to star mass ratio is smaller than a critical value
\begin{equation}
    \epsilon_{\rm p} < \epsilon_{\rm p,crit} = 1.32\frac{(1+\zeta)^2}{\left(k(1+\zeta) + \frac{\tilde{\tau}_e}{\tilde{\tau}_{a,e}}\right)^{3/2}}\frac{ \tilde{\tau}_e}{\tilde{\tau}_{a,e}}\left(\frac{\tilde{\tau}_e}{\tilde{\tau}_a}\right)^{3/2},
    \label{eq:epscrit}
\end{equation}
where $\zeta=m_1/m_2$ is the planet mass ratio, and, $\tilde{\tau}_e^{-1} = \tau_{e,1}^{-1}+\zeta\tau_{e,2}^{-1}$ and $\tilde\tau_{a,e}^{-1} \simeq \tau_{e,1}^{-1}-\zeta^2\tau_{e,2}^{-1} $ are related to the eccentricity damping timescales \citep[Eqs. 9 and 14]{Deck2015}.
Using the eccentricity damping expressions from type-I migration, we see that $\tilde{\tau}_e$ does not depends on $\zeta$ whereas the sign of $\tilde\tau_{a,e}$ is roughly prorptional to $1-\zeta$.
It is also straightforward to see that $\epsilon_{\rm p,crit}$ is independent of $\Sigma$.
In particular for $\zeta\geq1$ or $m_1 \geq m_2$, $\epsilon_{\rm p,crit}$ is negative and the resonance center is always stable\footnote{The discussion is slightly different for the 2:1 MMR, see \citet{Deck2015}.}.
In this Appendix, we highlight how our result would differ in the case of a more massive outer planet. 
Similarly to Section~\ref{subs:s0_ep}, we can restrict ourselves to the case of type-I migration by replacing the disk surface density by the rescaled surface density $\Siggap$ (Eq. \ref{eq:siggap}).
We still highlight with a dashed line the transition to the type-II migration regime.

\subsection{Fixed inner planet}\label{appsub:fixed}

\begin{figure}
    \centering
    \includegraphics[width = \linewidth]{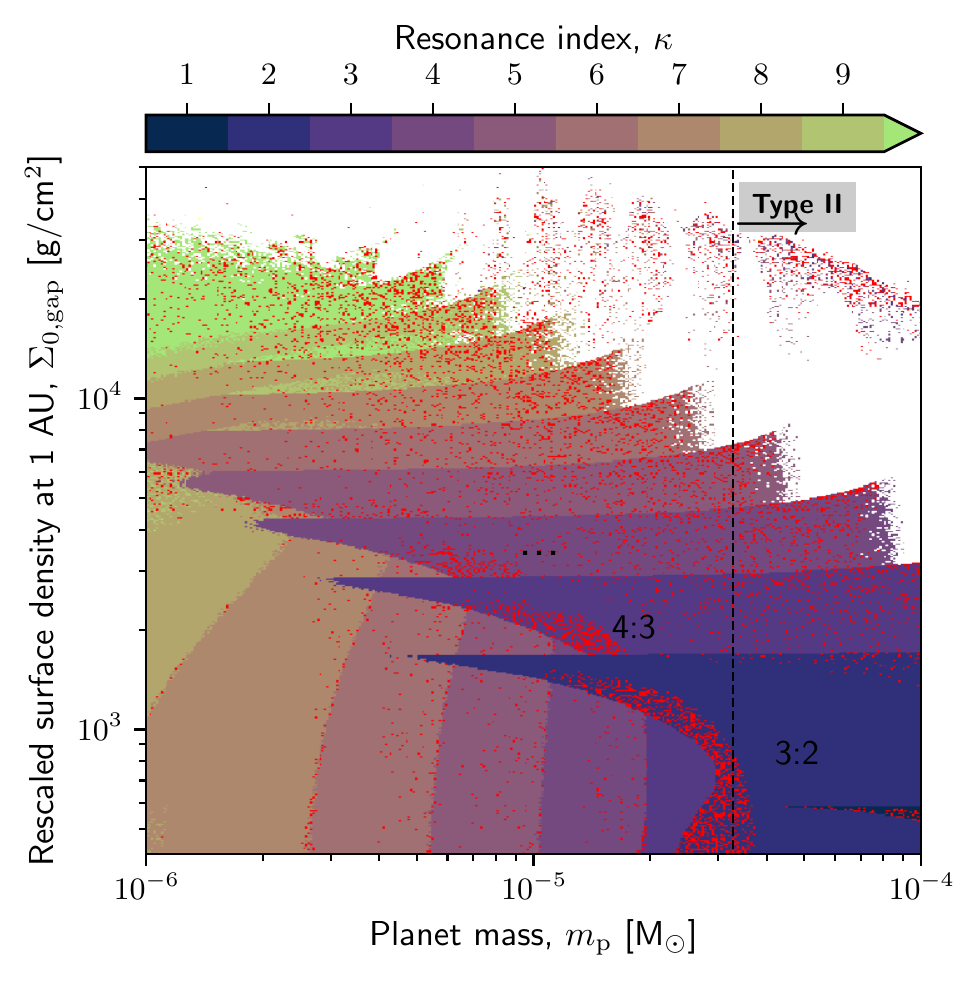}
    \caption{The final value of the continuous resonance index $\kappa$ as function of the rescaled surface density $\Siggap$ and $\mpl$ for systems with $\mpl= m_1m_2/2$. 
    The inner planet is fixed at the inner disk edge and the initial conditions are otherwise the same as in Figure \ref{fig:fiducial}. The black dashed line is the gap opening threshold. 
    Except for a few systems, planets with masses below $10^{-4}\Msun$ that would otherwise be captured in 2:1, do not stay there due to overstability.}
    \label{fig:m22m1}
\end{figure}

We simulate the same system conditions as in Subsection \ref{subs:s0_ep} with a more massive outer planet, $m_2=2m_1=2\mpl$, to see how the overstability impacts the final configuration at smaller planet-to-star mass ratio $\ep$. 
Figure \ref{fig:m22m1} shows the resulting 2D grid with varied $\Siggap$ and $\mpl$.
We observe a regime where the transitions between resonances do not depend on $\mpl$ at larger masses same as in Figure \ref{fig:fiducial}.
Unlike Figure \ref{fig:fiducial} we also see vertical transitions towards high $k$ for small $\mpl$ values, most likely due to overstability of the lower index resonances.
Indeed a more massive outer planet leads to overstable systems for small $\mpl$ even for $k>1$. 
According to \citet{Deck2015} the threshold value of $\ep = (m_1 + m_2)/M_*$, for overstability increases for all $k$ if the outer planet is more massive and generally decreases with increasing $k$. 
Moreover, we see that the systems close to the boundary to overstability are not in resonance according to Eq. \eqref{eq:delta}.
A more detailed study of these system showed that they experience a significant libration of their resonant angles, which is consistent with previous studies such as \citet{Xu2018}, that showed the continuity between the system escaping the resonance and systems with significant libration.
A difference in our results than from the Deck's theory is that $\epsilon_{\rm p,crit}$ seems to  be a decreasing function  of the surface density until it reaches the transition surface density observed in the equal mass case.
This more complex pattern was also observed by \cite{Xu2018}.

In the case of a more massive inner planet, the results would be qualitatively similar to the equal mass case detailed in Section \ref{sec:results}). 
Indeed, the overstability mainly occurs for $m_1\lesssim m_2$ if the eccentricity damping timescale is proportional to the planet masses \citep{Deck2015}.
In the context of the 2:1 MMR, the overstability can trigger for $m_1/12\lesssim m_2$.

\subsection{Both planets migrating}\label{appsub:mig}
\begin{figure}
    \centering
    \includegraphics[width = \linewidth]{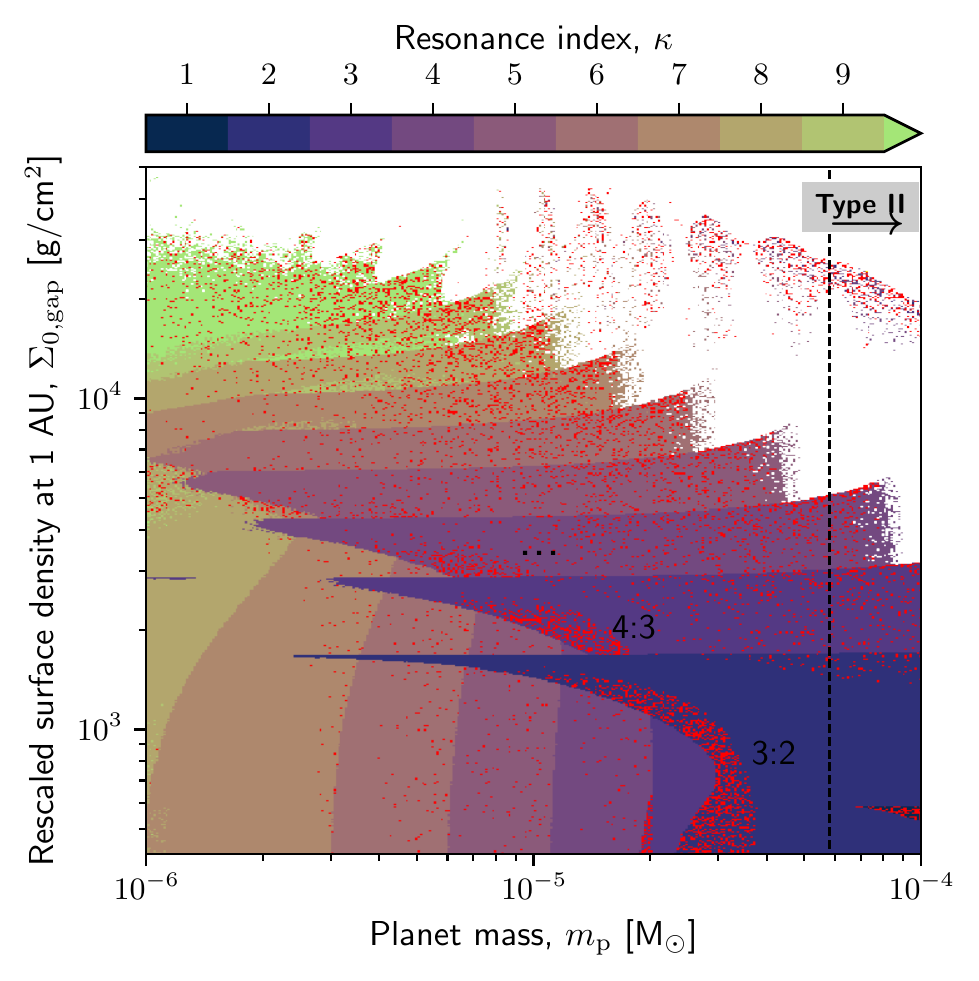}
    \caption{Same as Figure \ref{fig:m22m1} but here both planets migrate towards the inner disk edge from the positions $[0.5, 0.85]\, {\rm AU}$.}
    \label{fig:mig}
\end{figure}
We also investigate whether the qualitative behaviour is changed, if at all, when both planets in the simple two-planet system migrate towards the inner disk edge. 
The initial conditions of this run are the same as in Subsection \ref{appsub:fixed}, but now the inner planet migrates from $0.5\, {\rm AU}$ and the outer from  $0.85\, {\rm AU}$. 
As capture in MMR is only possible for convergent migration, we need to be in a configuration such that the outer planet is more massive and thus migrates faster than the inner one.
We use the same mass ratio, $m_2=2m_1=2\mpl$ as in Appendix \ref{appsub:fixed}. 
As the planet start further away in the disk, the transition towards the type-II regime occurs at a higher mass value.

Figure \ref{fig:mig} shows a 2D grid of final $\kappa$ values as function of $\Siggap$ and $\mpl$. 
We observe that the behaviour of the systems is similar to the precedent case.
Both the surface density transitions as well as the overstable region are similar.
Thus, the capture in resonance is not strongly affected by whether both planets are migrating or not, and only weakly by the period of the inner planet at capture, which changes the local surface density and scale height.
This figure also compares well with Figure \ref{fig:fiducial} in the stable regime.
These figures, \ref{fig:mig} and \ref{fig:m22m1}, show that overstability exists only for small $\mpl$ values and is present for higher $k$ MMRs if the outer planet is more massive, as predicted by \citet{Deck2015,Xu2018}. 
Qualitatively, the shape of the boundary is the same as seen in Figure 2 in \citet{Deck2015} and Figure 3 in \citet{Xu2018}.

\section{Comparing respectively the inner and outer planet pairs in the three-planet chain with the two-planet chain}\label{app:comp2v3}

\begin{figure}
    \centering
    \includegraphics[width = \linewidth]{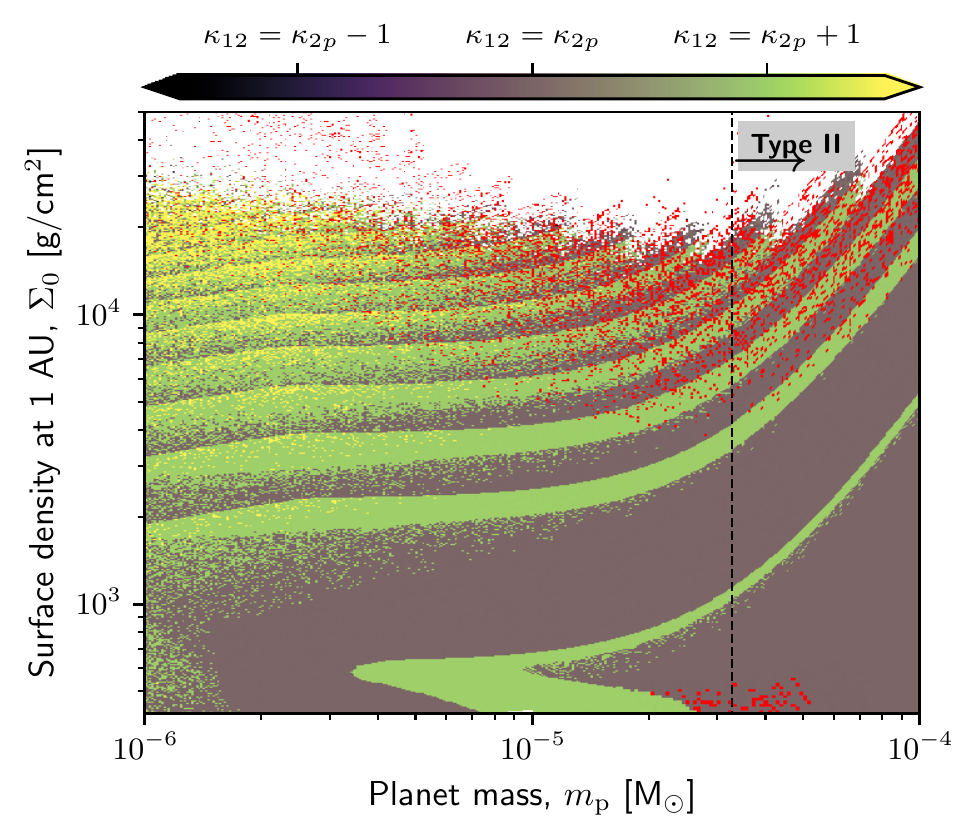}
    \caption{Difference between the resonant index $\kappa_{12}$ of the inner planet pair in a three-planet system with the index $\kappa_{2\mathrm{p}}$ of a two-planet system for the same planet masses and disk properties (the resonant indexes of the two- and three-planet systems are respectively displayed on Figures. \ref{fig:fiducial} and \ref{fig:3p}a). 
    See Figure \ref{fig:fiducial} for a full description of the setup. 
    As in previous figures, the red dots correspond to systems far from resonance. The black dashed line indicates the threshold for gap opening, $3.3\times 10^{-5}$M$_\odot$.}
    \label{fig:k12-k2p}
\end{figure}

\begin{figure}
    \centering
    \includegraphics[width = \linewidth]{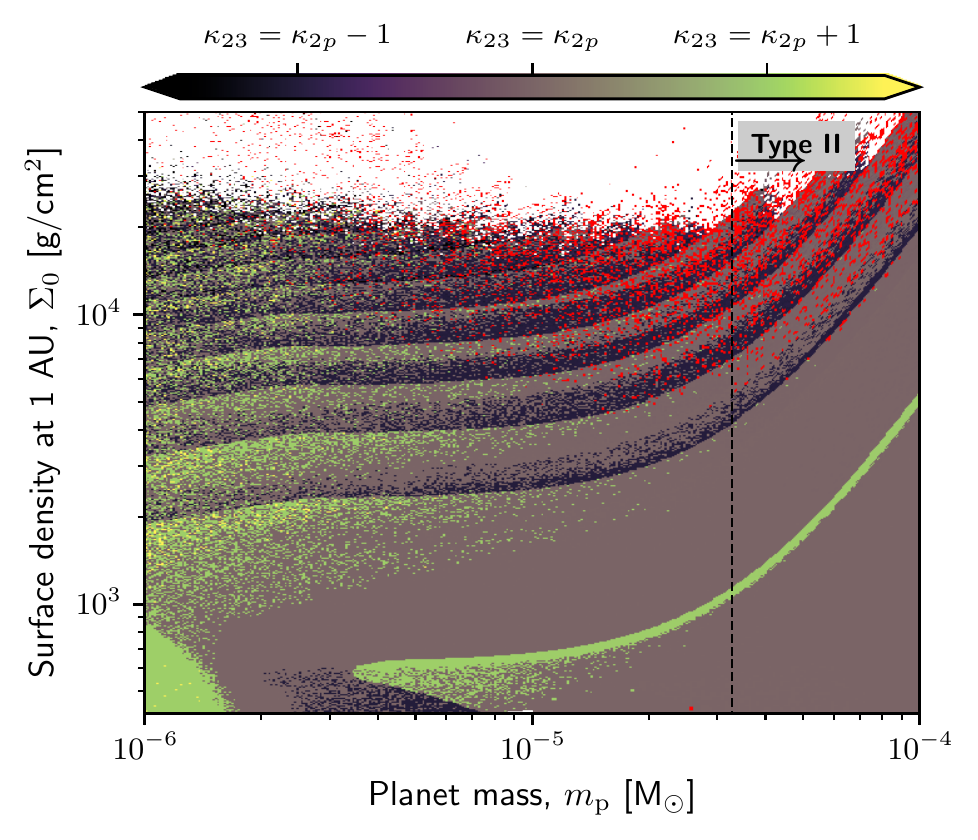}
    \caption{Same as Figure \ref{fig:k12-k2p} for the difference between the outer pair resonant index $\kappa_{23}$ (shown on Figure \ref{fig:3p}b) and the two-planet system resonant index $\kappa_{2\mathrm{p}}$.}
    \label{fig:k23-k2p}
\end{figure}

In section \ref{sec:3p}, we discuss the outcome of the resonant capture in the three-planet case with respect to the two-planet case. 
We show in Figure \ref{fig:k12-k2p} (respectively Figure \ref{fig:k23-k2p}), the difference between the resonant index of the inner (resp. outer) pair of a three-planet system and the resonant index of a two-planet system $\kappa_{12}-\kappa_{2\mathrm{p}}$ (resp. $\kappa_{23}-\kappa_{2\mathrm{p}}$) for the same planet mass and disk parameters.
As said in section \ref{sec:3p}, the inner pair of the three planet system can be pushed into a tighter resonance (higher index) by the capture of the outermost planet.
Before the capture of the outer planet, the inner pair has the same behaviour as a two-planet system.
For the outer pair, we note that the transitions are noisier, with pairs on both tighter and wider resonances (higher or lower indexes) than their two-planet counterparts.
\end{appendix}
\end{document}